\documentclass[11pt]{article}

\usepackage{amssymb,amstext,amsmath,amsthm}
\usepackage[dvips]{graphicx}
\usepackage{latexsym}
\usepackage{psfrag}
\usepackage{amsfonts}
\usepackage{bbm}
\usepackage{color}
\usepackage{footnpag} 

\usepackage{amsmath}
\usepackage{amsfonts}
\usepackage{graphicx}
\usepackage{psfrag}
\usepackage{array}
\usepackage{bbm}
\usepackage{hyperref}
\usepackage{amsthm}
\theoremstyle{definition}

\newcommand\beq{\begin{equation}}
\newcommand\eeq{\end{equation}}
\newcommand\be{\begin{eqnarray}}
\newcommand\ee{\end{eqnarray}}
\newcommand\beqa{\begin{eqnarray}}
\newcommand\eeqa{\end{eqnarray}}
\newcommand\bean{\begin{eqnarray*}}
\newcommand\eean{\end{eqnarray*}}
\newcommand{\bes}{\begin{eqnarray}}
\newcommand{\ees}{\end{eqnarray}}

\newcommand{\cC}{{\mathcal C}}
\newcommand{\cD}{{\mathcal D}}

\newcommand{\cG}{{\mathcal G}}

\newcommand{\cI}{{\mathcal I}}

\newcommand{\cM}{{\mathcal M}}

\newcommand{\cS}{{\mathcal S}}

\newcommand\R{{\mathbb R}}
\newcommand\Z{{\mathbb Z}}

\newcommand{\SO}{{\rm SO}}
\newcommand{\so}{\mathfrak{so}}

\newcommand{\su}{\mathfrak{su}}
\newcommand{\SU}{{\rm SU}}

\def\inv{{\mbox{\tiny -1}}}
\def\minus{{\mbox{\small -}}}
\def\plus{{\mbox{\tiny +}}}

\newcounter{letter} \newcounter{numeral} \newcounter{Numeral}

\newcommand\Tr{\mathrm{Tr}}
\def\extd{\mathrm {d}}
\def\vphi{\varphi}
\def\vphihat{\widehat{\varphi}}

\newcommand\e{{\mbox{e}}}

\newcommand\acts\triangleright

\begin{document}

%

%

\title{\large 
\bf Group field theory and simplicial gravity path integrals:\\
A model for Holst-Plebanski gravity}

\author{
Aristide Baratin${}^{a}$ and Daniele Oriti${}^{b}$
\medskip \\
{\small ${}^a$ 
\emph{Centre de Physique Th\'eorique, CNRS UMR 7644, Ecole Polytechnique}} \\
{\small 
\emph{F-9112 Palaiseau Cedex, France}}
\\
{\small ${}^b$ 
\emph{Max Planck Institute for Gravitational Physics, Albert Einstein Institute}} \\
{\small 
\emph{Am M\"uhlenberg 1,14467 Golm, Germany}}
}




\maketitle



\begin{abstract}
\noindent In a recent work, a dual formulation of group field theories as non-commutative quantum field theories has been proposed, 
providing an exact duality between spin foam models and non-commutative simplicial path integrals for constrained BF theories.   
In light of this new framework, we define a model for 4d gravity which includes the Immirzi parameter $\gamma$.  
It reproduces the Barrett-Crane amplitudes when $\gamma\!=\!\infty$, but differs from existing models otherwise; in particular it does not require  any rationality condition for $\gamma$.
We formulate the amplitudes both as BF simplicial path integrals with explicit non-commutative $B$ variables, and in spin foam form in terms of Wigner 15j-symbols. 
Finally, we briefly discuss the correlation between neighboring simplices, often argued to be a problematic feature, for example, in the Barrett-Crane model.
\end{abstract}

\bigskip \bigskip

%
\section{Introduction}
%

\noindent Group field theories \cite{gft} are quantum field theories showing up as a higher dimensional generalization of matrix models 
in background independent approaches to quantum gravity \cite{libro}.  The perturbative Feynman expansion generates stranded graphs dual to simplicial complexes of all topologies, weighted by spin foam amplitudes \cite{SF}. 
Conversely, it can be shown \cite{mikecarlo} that any spin foam model  admits a GFT formulation, which removes its dependence on the triangulation.  

Most spin foam and GFT models for quantum gravity are based on modifications of the Ooguri model for 4d 
$BF$ theory\footnote{with some exceptions, see for e.g  \cite{cubulation}.}.   
This approach is motivated by the fact that classical 4d gravity  can be expressed as a constrained $BF$ theory (Plebanski formulation) \cite{mike, DP-F} 
\beq \label{pleb}
S(\omega,B, \lambda)=\int_{\mathcal{M}} \Tr \, B \wedge F (\omega) + \lambda \, \cC(B), 
\eeq
for $\so(4)$ valued\footnote{we will restrict to the Riemanian signature in this paper.} 1-form connection $\omega$ and 2-form field $B$,  where $\cC(B)$ are polynomial (so-called simplicity) constraints  and $\lambda$ is 
some  Lagrange multiplier. The variation with respect to the Lagrange multiplier constrains $B$ to be 
a function of a tetrad 1-form field 
$B\!=\!\ast (e\wedge e)$, turning  BF to the Palatini action for gravity in the first order formalism. 
The Immirzi parameter $\gamma$, which plays a crucial role in loop gravity, can be introduced by replacing $B \!\to\! B + \frac{1}{\gamma}\!\ast\! B$ in the BF term of the action. 
Solving the constraints reproduces the Holst action \cite{Holst},  classically equivalent to Palatini gravity and starting point for the quantization leading to loop quantum gravity.

The spin foam quantization stems from a discretization of the classical theory, by choosing a triangulation $\Delta$ on $\cM$. 
While the most direct route to quantization would be to include a discrete analogue of the constraints  $\cC(B)$ into the definition of the measure of the discretized path integral \cite{reisenberger, kirilllaurent}
\beq \label{pathintegral}
I_{\Delta} = \int \cD[\omega_{\Delta}, B_{\Delta}]\delta(\cC(B_{\Delta})) \, \e^{i \Tr \, B_{\Delta}F_{\Delta}},
\eeq
the standard spin foam strategy consists of quantizing first the topological BF part of the discretized theory: the discretization and quantization of BF theories in any dimension are in fact well-understood \cite{kirilllaurent, Ooguri}.
The task is then to  implement a quantum version of the constraints in order to recover the gravity degrees of freedom.  
This has shown to be a quite subtle task, partly because  
of  the very simplicial setting in which the construction takes place: in fact, no standard canonical quantization procedure exists in such discrete setting, and of course things only become more difficult when the classical system to be quantized is a background independent simplicial gravity theory\footnote{see however \cite{BiancaPhilip} for a recent proposal  of a general canonical formalism for simplicial gravity.}. 
Proposals for the implementation of the constraints motivated by the geometric quantization of simplicial structures \cite{barbieri, baezbarrett} first led to the famous Barrett-Crane model \cite{BC}, and more recently to the  EPRL model \cite{EPRandL}, which includes the Immirzi parameter and reduces to  Barrett-Crane when $\gamma \!=\!\infty$. 

One of the main difficulties that this strategy encounters stems from the non-commutative nature of the geometrical variables in the BF Ooguri model \cite{barbieri} -- 
beginning with the quantum $B$ variables themselves represented as generators of the gauge group -- which  
obscures the geometrical interpretation of the constraints. As an attempt to remedy this problem, Livine and Speziale suggested \cite{LivineSpeziale} to rewrite the BF amplitudes in a basis of Perelomov group coherent states and to interpret the coherent state labels as classical  bivectors (though with quantized norm) on which to impose the constraints.  The realization of this idea led to the FK$_{\gamma}$ model \cite{FK} which, remarkably, coincides with EPRL when $\gamma \!<\!1$.  
Using the coherent sate representation, it can be shown that both models exhibit a path integral-like formulation \cite{FKpathintegral}. 
Though it provides a powerful tool to relate the models to Regge gravity in the semi-classical limit \cite{asymptotics}, this formulation, 
which involves a quite specific and non-standard action, 
is however far remote from the original  path integral (\ref{pathintegral}).

The goal of the series of papers \cite{aristidedaniele, danielearistideBC} and the present work, in the spirit of  earlier works \cite{eteravalentin}  by Bonzom and Livine, 
is instead to dig deeper into the relation between spin foam models and simplicial path integrals of the type (\ref{pathintegral}). 
Our main result is to show that generic spin foam models based on quantum BF theory have a dual formulation as a version of (\ref{pathintegral}) in which functionals of the discrete bivectors $B_{\Delta}$ are endowed with a non-commutative structure (star product) deforming the usual point-wise product. 
Such a formulation of the path integral\footnote{this formulation is adapted to the quantization of classical systems with a `curved' phase space, see \cite{danielematti} for the simple example of a quantum system on the group $\SO(3)$.}, in contrast to standard constructions with commutative variables \cite{thomasmuxin}, 
captures the key aspect of non-commutativity of bivectors in spin foam models, 
covariant counterpart of the non-commutativity of the flux variables in loop  gravity \cite{ACZ, flux}.   
It is important to note that this non-commutativity is not an anomaly of quantization, as it can be traced back to the classical theory \cite{ACZ, twisted, laurentmarc, biancajimmy}. 

This result is important for the study of spin foam models, in many respects. 
It gives a direct correspondence between purely algebraic amplitudes describing the quantum geometry on one hand, and a measure on the  variables of the classical theory on the other. It allows for a direct comparison between the spin foam quantization and a proper path integral quantization of Holst-Plebanski gravity \cite{Plebcanonical, sergei}. It also opens the way for a precise study of the consequences of non-commutativity of the  geometry  inherent to these models. 

\

The duality between spin foam models and the  path integrals (\ref{pathintegral}) is realized at the level of the generating group field theories. 
We will thus work in this very general setting, though the construction could also be carried out directly at the level of the amplitudes. 
The mechanism is the following. As we will review in Section \ref{BF},  in addition to its usual formulations in terms of gauge invariant group fields  $\vphi(g_1, \cdots g_4)$ or of its Peter-Weyl tensor components, 
the Ooguri GFT model for $\SO(4)$ BF theory has a dual formulation in terms of fields on four copies of the Lie algebra 
$\so(4)$, obtained by a  Fourier transform: 
\beq
\vphihat(x_1,\cdots x_4)\!:=\! \int[\extd g_i]^4\, \varphi(g_1, \cdots g_4)\, \e^{i \Tr x_1 g_1} \cdots \e^{i \Tr x_4 g_4}
\eeq
which endows the space of fields  with a non-commutative $\star$-product (dual to group convolution). 
It can be shown that the gauge invariance of field translates into a closure condition   $x_1 + \cdots + x_4 =0$ for the $\so(4)$ variables, which have a direct interpretation as the discrete $B$ variables labeling the faces of a tetrahedron. In fact, in this representation, the GFT Feynman amplitudes are simplicial BF  path integrals \cite{aristidedaniele}. 
In such a representation, where all the geometrical variables are explicit, 
constrained models for gravity take a very suggestive form in terms of constrained  fields $(\cS\star \vphihat)(x_j)$ for some  
functions $\cS(x_j)$ constraining the bivectors. By construction, the Feynman amplitudes are simplicial path integrals for constrained BF theories.

In principle, every GFT and spin foam model for gravity based on BF theory, thus including EPRL/FK, can be formulated this way, with more or less natural forms for the constraint functions $\cS$. 
In this paper, however,  we rather follow a constructive approach: in Section \ref{geomodels}, we {\sl define} $\cS$ in  the most natural way in this framework:  namely in terms (non-commutative) Dirac distributions $\cS(x_j) \!=\! \delta(\cC(x_j))$ where  $\cC(x_j)$ are the discrete simplicity constraints. 
The use of Dirac distributions effectively amounts to constraining the measure on the bivector variables. 
As we will see, the fact that the constraints are imposed on the group field, hence in all tetrahedra, will automatically lead to additional constraints on the connection in the path integral form of the amplitudes \cite{sergei}. 

We will work  with the linear form of the discrete simplicity constraints \cite{serguei1, EPRandL, FK, danielesteffen}, with Immirzi parameter,
and a minimal extension of the group field formalism to include the normals to tetrahedra as an additional variable of the field: 
this allows us to implement the constraints in a gauge covariant way. 
In this formalism, polynomial boundary observables are labelled by so-called projected spin-networks \cite{projected}. 
We thus obtain a constrained GFT formulated as a theory of dynamical (non-commutative) geometric tetrahedra, which interact in the simplest possible way, as dictated by the star product. Its Feynman amplitudes define a spin foam model for gravity with Immirzi parameter $\gamma$, which gives a variant the Barrett-Crane model when $\gamma\!=\!\infty$ but differs from the existing models for generic values of $\gamma$. In particular it does not require any rationality condition for $\gamma$. 
This  model is formulated both as a path integral (\ref{pathintegral}) and in terms of Wigner 15j-symbols, in Equ. (\ref{explPI}) and (\ref{spinfoam}) below: 
this is the main result of the paper. 

The framework will also allow us to take a new (covariant) look at the peculiar features of the path integral amplitudes induced by non-commutativity.    
In particular we briefly discuss Section \ref{ultralocality} how the so-called `ultralocality problem' anticipated for the Barrett-Crane model manifests itself in our framework and argue that it may disappear in a suitable semi-classical limit involving a commutative limit. 

In Section \ref{concl} we conclude and sketch some directions for future work.  


\section{GFT models for BF theory} 
\label{BF}

In this section, we start by recalling the standard Ooguri GFT for BF theory and its non-commutative bivector representation. 
We then present an extension of the GFT formalism, where the usual field variables, associated to the four triangles of a tetrahedron,  
are supplemented by an $S^3$ vector playing the role of the normal to the tetrahedron. 
As we will see, it will allow us to implement the linear simplicity  constraints (\ref{simp}) in a covariant way.

Our notations and conventions are as follows.  We identify functions on $\SO(4)$ with functions on $\SU(2)^\minus\!\times\! \SU(2)^\plus / \Z_2$ and denote by $g\!=\!(g^\minus, g^\plus)$  the $\SU(2)$ decomposition of the field variables.  We also use the decomposition of $\so(4)$ in anti-sef dual and self dual sectors $\so(4)\!=\!\su(2)^\plus \oplus \su(2)^\minus$ and denote by $x\!=\!(x^\minus, x^\plus)$ the corresponding decomposition of its elements. 
From Sec. \ref{bivrep} on, based on the $\SO(3)$ Fourier transform\footnote{An extension of the group transform to the whole $\SU(2)$ has been developed in \cite{karim}. 
We do not use it in this work, because on the one hand we do not expect the results to be very much modified by such extension, and on the other hand the general case would entail a more involved notation. Note also that a different $\SU(2)$ transform has been proposed and studied in \cite{Voros}.} \cite{PR3}, 
we further assume an invariance of group functions under $g \to -g$, so that they are effectively functions
on $\SO(3)\! \times\! \SO(3)$. 

\subsection{Connection and spin formulations} 

\begin{figure}[h]
\includegraphics[scale=.4]{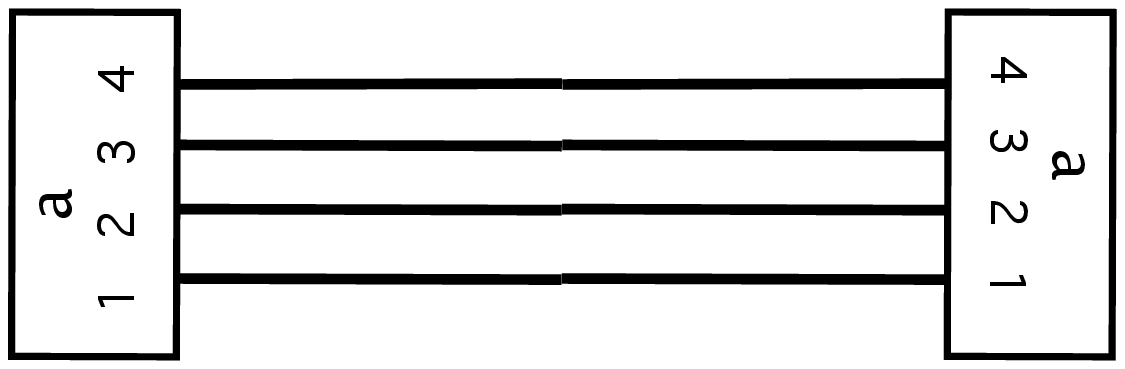}
\includegraphics[scale=.4]{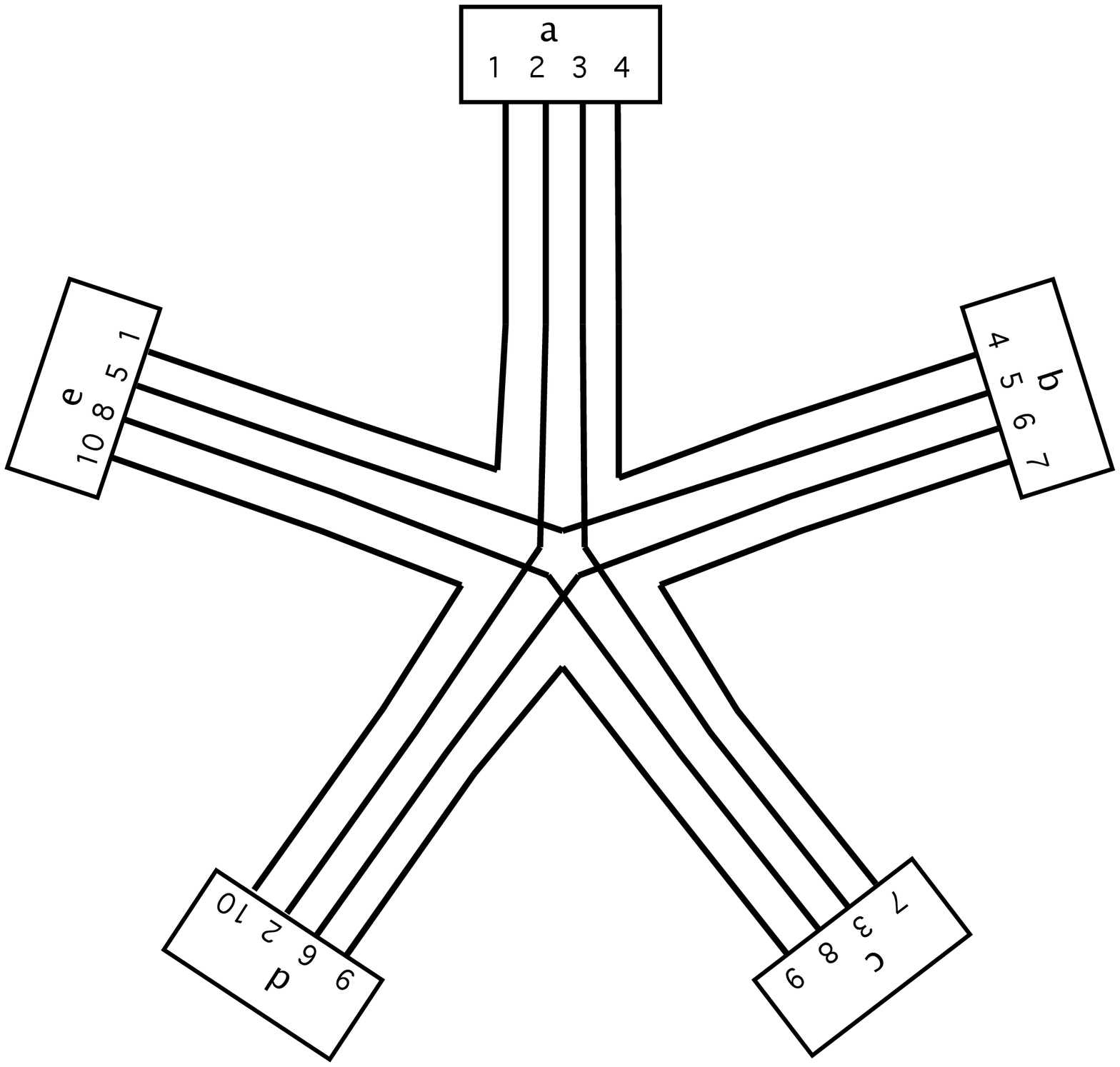}  
\caption{GFT propagator and vertex}
\label{gftfig}
\end{figure}

\noindent In the standard  connection formulation, the Ooguri GFT model \cite{Ooguri} for BF theory is described in terms of a field  $\vphi(g_1, \cdots g_4)$ on four copies of the gauge group,  satisfying the gauge invariance condition: 
\beq \label{gauge}
\forall  h \in \SO(4), \quad \vphi(g_1,\cdots  g_4) = \vphi(hg_1, \cdots hg_4)
\eeq
The dynamics is governed by the action: 
\beq \label{actiongroup}
S = \frac{1}{2} \int [\extd g_i]^4 \varphi^2_{1234}  + \frac{\lambda}{5!}  \int [\extd g_i]^{10}  
\varphi_{1234} \,\varphi_{4567} \,\varphi_{7389}\, \varphi_{962\,10} \,\varphi_{10\,851} 
\eeq
where $\vphi_{1234}$ is a shorthand notation for $\vphi(g_1, \cdots g_4)$, $\extd g$ is the normalized Haar measure and $\lambda$ is a coupling constant. 
The perturbative expansion in $\lambda$ generates 4-stranded graphs dual to 4d simplicial complexes (see Fig \ref{gftfig}):
if one associates the field variables to the four triangles of a tetrahedron, the quintic interaction sticks five tetrahedra together a common triangle to form a 4-simplex;  the kinetic term dictates the gluing rules for 4-simplices along tetrahedra.  



By using the harmonic analysis on $\SO(4)$, the gauge invariant field is expanded into four $\SO(4)$ irreducible representations, 
labelled  by pairs of  $\SU(2)$ spins $J\!=\!(j^\minus,j^\plus)$, and  4-valent intertwiners $\imath\!=\!(\imath^\minus, \imath^\plus)$ labelled by a pair of intermediate $\SU(2)$ spins.  
The interaction vertex is  expressed in terms of the $\SU(2)$ Wigner 15j-symbols \cite{Ooguri}. 
In this formulation, the Feynman amplitudes takes the form of a state sum model on the simplicial complex dual $\Delta$ to the graph:
\be \label{ooguri}
I_{\Delta} =\, \sum_{\{J_t, \imath_\tau\}} \prod_t d_{j_t^\minus} d_{j_t^\plus} \prod_ \sigma\, \big \{ 15j \big\}^\minus_\sigma \big\{ 15j \big\}^\plus_{\sigma} 
\ee
The sum is over the $\SO(4)$ representations $J_t$ and intertwiners $\imath_{\tau}$ labeling the triangles and the tetrahedra; $d_j \!=\! 2j+1$ is the dimension of the $\SU(2)$ representation $j$. The symbol for the vertex amplitude (associated to each 4-simplex), is the product of  two $\SU(2)$  Wigner $15j$-symbols.

\subsection{Non-commutative Fourier transform and bivector formulation}
\label{bivrep}

The simplicial geometry encoded in the model (\ref{actiongroup}) is best understood in a dual formulation, coined `metric representation' in \cite{aristidedaniele}, 
obtained by a group Fourier transform of the field. The relevant Fourier transform here is the obvious extension of the non-commutative $\SO(3)$ Fourier transform \cite{PR3,karim,laurentmajid} 
to the group $[\SO(3)\times \SO(3)]^4$: 
\beq
\vphihat(x_1,\cdots x_4)\!:=\! \int[\extd g_i]^4\, \varphi(g_1, \cdots g_4)\, \e^{i \Tr x_1 g_1} \cdots \e^{i \Tr x_4 g_4}
\eeq
The variables $x_i$ belong to the Lie algebra $\so(4)\!=\!\su(2) \oplus \su(2)$. 
The kernel of the Fourier transform is a product of `plane waves' 
$E_g(x) =\e^{i\Tr xg}$, 
where the trace $\Tr$ is defined in terms of the usual trace of $2\times 2$ matrices\footnote{Let $\tau_j$ be i times the Pauli matrices, then $\mbox{tr} \tau_i \tau_j \!=\! - \delta_{ij}$. Given and $\SU(2)$ element $u\!=\!\e^{\theta n^j \tau_j}$ parametrized  by the angle $\theta \in [0,\pi]$ and the unit $\R^3$-vector $\vec{n}$ and $a\!=\!a^j\tau_j$ in the algebra $\su(2)$, we thus have $\mbox{tr}[a u] \!=\! - \sin \theta \vec{n} \cdot \vec{a}$.  
Also $\epsilon_u \!:=\!\mbox{sign} (\mbox{tr} u) \!=\!  \mbox{sign} (\cos\theta)$.
}
as $\Tr xg \!=\! \sum_{\pm} \epsilon_{g^\pm} \mbox{tr}[x^\pm g^\pm]$ with  $\epsilon_{g^\pm}\!=\!\mbox{sign} (\mbox{tr} g^\pm)$.  
Thus $E_g(x)$ is itself a product  of two $\SO(3)$ plane waves  
$\e_{g^\pm}(x^\pm)\! :=\!\e^{i\epsilon_{g^\pm}\mbox{\footnotesize tr}x^\pm g^\pm}$. 
The plane waves satisfy the properties:
\beq \label{prop}
\int \extd^6 x \, E_g(x) = \delta(g), \quad E_{g^\inv}(x) = E_g(-x)
\eeq
Here $\extd^6 x$ is the Lebesgue measure on $\so(4)\!\sim\! \R^6$ and 
$\delta(g)\!:=\!\delta_{\SO(3)}(g^\plus)\delta_{\SO(3)}(g^\minus)$ 
acts as the delta distribution on group fields.  We deduce from these the following expression of the GFT action (\ref{actiongroup})  in terms of the dual field $\vphihat$:
\beqa \label{actionbiv}
S \!\!&=&\!\!\frac{1}{2} \int [\extd^6 x_i]^4 \vphihat_{1234} \star \vphihat_{\minus 1 \minus 2 \minus 3\minus 4} \nonumber  \\
&& \hspace{1cm}  + \frac{\lambda}{5!}  \int [\extd^6 x_i]^{10}  \,
\vphihat_{1234} \star \vphihat_{\minus 4567}\star  \vphihat_{\minus 7\minus 389}\star  \vphihat_{\minus 9\minus 6\minus 2\,10} \star \vphihat_{\minus 10\,\minus 8\minus 5\minus 1} 
\eeqa
Notations are as follows. $\vphihat_{\pm 1 \pm 2\pm 3\pm4}$ is a shorthand notation for $\vphihat(\pm x_1, \cdots, \pm x_4)$. 
The $\star$-product is defined on $\SO(3)$ plane waves as $\e_g\star \e_{g'}(x)\!=\!e_{gg'}$, extended to 
$E_g \star E_{g'}(x) \!=\! E_{gg'}(x)$ and  by linearity to the algebra functions. 
In the  expression above, it is understood that the $\star$-product pairs repeated indices: for example,  the first product of the interaction term 
is a product of functions of the variable $x_4$, $\vphihat_{1234} \star_{x_4} \vphihat_{\minus 4567}$.
To recover (\ref{actiongroup}) from (\ref{actionbiv}), one expands the dual fields in group modes; 
the integration over the variables $x_i$ produces delta functions $\delta(g^{\minus 1}_i g'_i)\!\!=\!\!\int \extd x_i E_{g^\inv_ig'_i}(x_i)$ which identify the group elements associated to the same index.  

Gauge invariance (\ref{gauge}) translates into the invariance of the dual field under $\star$-multiplication by a product of four plane waves 
$E_h(x_1)\cdots E_h(x_4)\!=\!E_h(x_1+..+x_4)$ labelled by the same $h$:
\beq \label{gaugebiv}
\forall  h \in \SO(4), \quad \vphihat= E_h \cdots E_h \star \vphihat
\eeq
Integrating over $h$ on both sides of this equality gives: 
\beq  \label{gaugeproj}
\vphihat = \delta(x_1\!+ ..+\!x_4) \star \vphihat, \qquad \delta(x):=\int \extd h E_h(x)
\eeq
where $\delta$ plays the role of non-commutative delta function on algebra functions $\delta\star \phi(x) \!=\!\phi(0) \delta(x)$. 
In words, gauge invariance corresponds to a constraint  on the dual fields imposing the closure $x_1+..+x_4=0$ of its variables\footnote{In terms of the canonical theory, this is indeed just the standard Gauss law corresponding to $\SO(4)$ gauge invariance in flux variables \cite{flux}}. 
It is interesting to note that, in the non-commutative setting,  the closure constraint is implemented by a projector, since 
$\delta\star \delta\!=\! \delta$.  Geometrically, $\vphihat$ represents a tetrahedron whose four faces are labeled by a bivector $x_i^{IJ}$. 
The glueing rules for tetrahedra dictated by the action  (\ref{actionbiv}) corresponds to the identification of the face bivectors, 
modulo a sign encoding  a flip of the face orientation.

More precisely, propagator and vertex in this representation are given by
\beq \label{Feynmanfunctions2}
P(x, x') = \prod_{i=1}^4 \delta_{\minus x_i}(x'_i), \qquad 
V(x, x') = \int  [\extd h_{\ell}]^5 \, \prod_{i=1}^{10} ( \delta_{\minus x^{\ell}_i} \star\, E_{h_\ell h^{\minus 1}_{\ell'}})(x^{\ell'}_i)
\eeq
where $i$ labels the oriented strands (triangles) and $\ell$ the half lines (tetrahedra) of the graphs in Fig \ref{gftfig} and
$\delta_{x}(y)\!:=\!\delta(x-y)$, with $\delta$ defined as in  (\ref{gaugeproj}). 
In terms of simplicial geometry, the vertex function encodes the identification, for each of the ten triangles $i$ of a 4-simplex, of the two bivectors 
$x^\ell_i, x^{\ell'}_i$ associated to it, corresponding to the two tetrahedra  $\ell, \ell'$ sharing the triangle,  up to parallel transport $h_\ell h^{-1}_{\ell'}$ 
from one tetrahedron to another \cite{aristidedaniele}. 
The sign difference reflects the fact, in an oriented 4-simplex, a triangle inherit opposite orientations from the two tetrahedra sharing it. 
The integration over $h_\ell$ implements the gauge invariance (\ref{gaugebiv}). 
We have chosen to gauge-average the vertex here, but since gauge averaging is a projection $\widehat{C}^2  \!=\! \widehat{C}$,  
one could instead gauge-average the propagator or both vertex and propagator without affecting the amplitudes.  

The Feynman amplitudes are obtained by taking the $\star$-product of propagator and vertex functions, following the strands of the graph \cite{aristidedaniele}.
The structure of the $\star$-product gives a clear geometrical meaning to the algebraic expressions. 
In particular, commutation with a plane wave signifies a change of frame: $E_h\star\vphi\! =\! \vphi^h\star E_h$, with $\vphi^h(x) \!=\! \vphi(h^{\minus 1}xh)$.   

For a given closed graph dual to a simplicial complex,  this results in integrals over group elements $h_{\tau\sigma}$ labelled by adjacent pairs $\{$tetrahedron, 4-simplex$\}$, interpreted as parallel transport from the center of the 4-simplex $\sigma$ to the center of its boundary tetrahedron $\tau$,
and over $\so(4)$ variables $x_t^\tau, x_t^\sigma$, interpreted as the same bivector of $t$ seen in different frames associated to the tetrahedra and 4-simplices sharing $t$. In what follows we set $h_{\sigma\tau}\!:=\!h_{\tau\sigma}^{\minus 1}$ and denote by $h_{\tau\tau'}\!=\! h_{\tau\sigma} h_{\sigma \tau'}$ the holonomy 
between two neighboring tetrahedra through an adjacent 4-simplex.  
The integrand factorizes into contributions of each loop of strands of the graph, dual to a triangle, taking the form of 
a $\star$-product of delta functions identifying all variables $x_t^\tau, x_t^\sigma$ labelled but the  same $t$, up to parallel transport between the corresponding frames. 

After integration over all variables but one per triangle $x_t\!:=\!x_t^{\tau_0}$ associated to a `reference' tetrahedron $\tau_0(t)$, the amplitude reads: 
\beq \label{amplitudeBF}
\cI_{BF}\! =\! \int \! \prod_{<\tau\sigma>} \!\extd h_{\tau\sigma} \prod_t \extd^6 x_t \, e^{i \sum_t \Tr \, x_t H_t}
\eeq
where $H_t\!=\!h_{\tau_0\tau_1} \cdots h_{\tau_{N_t}\tau_0}$ is the holonomy along the  loop of $N_t+1$ tetrahedra sharing $t$ calculated for a choice of orientation and reference tetrahedron\footnote{The amplitude does not depend on these choices.}.
The integrand is the exponential of the discrete BF action, resulting from 
from the star product of $N_t$ plane waves for each $t$:
\beq
E_{h_{\tau_0 \tau_1}}\star \cdots \star E_{h_{\tau_{N_t} \tau_0}}(x_t) = e^{i\Tr \, x_t H_t}
\eeq
The GFT amplitudes in the bivector representation thus take the form of simplicial path integrals for BF theory,  where field variables $x \!\in\! \so(4)$ and group elements $h\!\in\! \SO(4)$ arising from gauge invariance play the respective roles of  discrete B field and discrete connection. 

The bivector formulation of GFT suggests clear routes for defining geometrical models, by means of constraints operators implementing  the simplicity constraint on the field variables. The linear constraints (\ref{simp}), however,  involve an another geometrical variable: the normal to the tetrahedron.  
In the next section, we review an extension of the usual GFT formalism introduced in \cite{danielearistideBC},  which include the normals as an additional field variable. 
Although the extended GFT generate the same BF amplitudes for closed graphs, it will allow us to impose the constraints covariantly, that is on fields that are gauge-invariant under a simultaneous $\SO(4)$ rotation of both bivectors and normal vector. 

\subsection{Introducing normals: extended GFT formalism} 

In the extended GFT formalism \cite{eteradaniele, danielearistideBC}, the basic group field $\vphi_k(g_1, \cdots g_4)$ is supplemented with a  fifth variable $k\!\in\! \SU(2)\!\sim\!\cS^3$, viewed as a unit vector  in $\R^4$. 
In geometrical models, $k$ will be interpreted as  the normal to a tetrahedron. 
Gauge invariance  (\ref{gauge}) is replaced by a gauge covariance  with respect to the normal $k$: 
\beq \label{extgauge}
\forall h, \quad \vphi_k(g_1, \cdots g_4)= \vphi_{h \acts k}(h g_1, \cdots h g_4) 
\eeq
where $h \acts k:=h^{\plus} k (h^\minus)^{\minus 1}$ is the normal rotated by $h$. 
Clearly, the field obtained by integrating over the normals obeys the gauge invariance  (\ref{gauge}). 

The dynamics is governed by the action:
\beq \label{extaction}
S[\vphi]  = \frac{1}{2}\int  [\extd g_i]^4 \extd k \, \varphi^2_{k1234}  + \frac{\lambda}{5!} \! \int [\extd g_i]^{10} [\extd k_i]^5 \, 
\varphi_{k_11234} \varphi_{k_24567} \varphi_{k_37389} \varphi_{k_4962\,10} \varphi_{k_510\,851} 
\eeq
where $\vphi_{k1234}$ is a shorthand notation for $\vphi_k(g_1, \cdots g_4)$, $\extd g$ and $\extd k$ are the Haar measures on $\SO(4)$  
and $\SU(2)$. Hence, whereas the interaction polynomial does not couple the normals,  the kinetic term, which encodes the glueing rule of 4-simplices along a tetrahedron, identifies both group elements and normals. It is  already clear from the structure of this action that the amplitudes of closed diagrams will not depend on the normals; the extended formulation only modifies the structure of boundary states. 

Note that gauge covariance (\ref{extgauge}) induces an invariance under the stabilizer group $\SO(3)_k\!=\!\{h\!\in\! \SO(4), h^\plus k (h^\minus)^{\minus 1} \!=\! k\}$ of the normal $k$, 
affecting only the four group arguments of the field. 
Upon Peter-Weyl decomposition, gauge invariant fields are expanded into four irreducible $\SO(4)$ representations (given by pairs of $\SU(2)$ spins $J_i\!=\! (j_i^\minus, j_i^\plus)$, $i=1\cdots 4$), each of which can be further decomposed into $\SO(3)_k$ representations. A set of basis functions is given by:
\beq \label{spinbasis}
\Psi^{(J_i, k_i, j)}_{m^\minus_i, m^\plus_i}(g_i;k) = \left(\prod_{i=1}^4 D^{j_i^\minus}_{n^\minus_i m^\minus_i}(g_i^\minus) 
D^{j_i^\plus}_{n_i^\plus m_i^\plus}(g_i^\plus) \widetilde{C}^{j_i^\minus j^\plus_i k_i}_{n^\minus_i n_i^\plus p_i}(k)\right)  (\iota_j)^{k_i}_{p_i}
\eeq
where repeated lower indices are  summed over. $D^{j^\pm}(g^\pm)$ are the $\SU(2)$ Wigner matrices, $(\iota_j)^{k_i}$ form a basis of four-valent $\SO(3)$ intertwiners, labelled by an intermediate spin $j$. 
The k-dependent coefficients, 
defined in terms of the 
$\SO(3)$ Clebsch-Gordan coefficients $C^{j^\minus j^\plus k}_{mnp}$ as:
\beq \label{Ctilde}
\widetilde{C}^{j_i^\minus j^\plus_i k_i}_{m^\minus_i m_i^\plus p_i}(k) = \sum_{m} C^{j_i^\minus j^\plus_i k_i}_{m m_i^\plus p_i} D^{j^\minus_i}_{m m^\minus_i}(k)
\eeq
define a tensor that intertwines  the action of $\SO(3)_k$ in the representation $j^\minus_i \otimes j^\plus_i$ and the action of $\SO(3)$ in the representation $k_i$.
Namely, given $\mathbf{u}_k \!=\! (k^{\minus 1} u k, u)\!\in\! \SO(3)_k$, we have:
\beq
\widetilde{C}^{j_i^\minus j^\plus_i k_i}_{m^\minus_i m_i^\plus p_i}(k) D^{j^\minus_i}_{m^\minus_i n^\minus_i}(\mathbf{u}_k^\minus) D^{j^\plus_i}_{m^\plus n^\plus_i}(\mathbf{u}_k^\plus)
= \widetilde{C}^{j_i^\minus j^\plus_i k_i}_{n^\minus_i n_i^\plus q_i}(k) D^{k_{i}}_{q_i p_i}(u).
\eeq
(\ref{spinbasis}) corresponds to  the vertex  structure of the so-called projected spin networks of the covariant approach to loop quantum gravity \cite{projected}, 
which thus define a basis for polynomial gauge invariant observables (and
thus boundary states) in the extended GFT formalism. 

Just as in the standard formulation, the bivector representation of the GFT is obtained by Fourier transform of the field
\beq
\vphihat_k(x_1,..,x_4)\!:=\! \int[\extd g]^4\, \varphi_k(g_1,..,g_4)\, \e^{i \Tr x_1 g_1} \cdots \e^{i \Tr x_4 g_4}
\eeq
Gauge invariance translates into:
\beq \label{extgaugebiv}
\forall h, \qquad \vphihat_k = E_h \cdots E_h \star \vphihat_{h^\inv \acts k}
\eeq
where $h^\inv \acts k\!=\!(h^{\plus})^{\minus 1}  k h^\minus$,  thus implemented by the gauge invariance projector acting on extended fields as
\beq \label{gaugeproj}
(\widehat{C} \acts \vphihat)_k = \int \extd h \, E_h \cdots E_h \star \vphihat_{h^\inv \acts k}
\eeq
Note that upon integration over the normal, gauge invariance gives the closure of the four bivector variables:
in fact if $\widehat{\psi} \!=\! \int \!\extd k \vphihat_k$, then
$\widehat{\psi}  = \delta(x_1 +.. + x_4)  \star \widehat{\psi}$. The action is the obvious extension of (\ref{actionbiv}). 
It is interesting to write the interaction in terms of $\widehat{\psi}$, to emphasize the fact it implements the closure constraints: 
\beqa \label{actionextbiv}
S \!\!&=&\!\!\frac{1}{2} \int [\extd^6 x_i]^4 [\extd k] \, \vphihat_{k1234} \star \vphihat_{k\minus 1 \minus 2 \minus 3\minus 4} \nonumber  \\
&& \hspace{1cm}  + \frac{\lambda}{5!}  \int [\extd^6 x_i]^{10}\,
\widehat{\psi}_{1234} \star \widehat{\psi}_{\minus 4567}\star  \widehat{\psi}_{\minus 7\minus 389}\star  \widehat{\psi}_{\minus 9\minus 6\minus 2\,10} \star \widehat{\psi}_{\minus 10\,\minus 8\minus 5\minus 1} 
\eeqa

The propagator of the extended GFT is supplemented with an additional strand which identifies the normals $k$ up to parallel transport arising from gauge invariance. 
Just as in the non-extended case, gauge invariance can be implemented in the vertex only, or in the propagator, or in both vertex and propagator. 
Choosing the first case, the propagator then reads, in the bivector formulation:
\beq  \label{propext}
P(x,k; x',k') = \prod_{i=1}^4 \delta_{\minus x_i}(x'_i) \, \delta(k' k^{\minus 1}) 
\eeq
The additional contribution reduces the number of $\SU(2)$ variables to one per link, hence, in terms of the dual simplicial complex, to one $k_{\tau}$ per tetrahedron. 
The integrals over the normals on the internal links (bulk tetrahedra) drop from the amplitudes; just like in the standard formulation, the extended GFT generate simplicial BF path integrals as Feynman amplitudes. However, due to the extension of the GFT field to include the normal vectors to tetrahedra, the boundary states appearing in the amplitudes for GFT n-point functions are differerent from those of the standard Ooguri model, recovered only after averaging out the normal variables independently at each tetrahedron in the boundary.

\section{Geometrical models} 
\label{geomodels}

We have seen in the previous section that, in the bivector representation of the GFT for BF theory,  
the field $\widehat{\psi} \!=\! \int \!\extd k \vphihat_k$ represents a tetrahedron characterized by four bivectors  $x_j$,  $j=1...4$  playing the role of discrete $B$ field; gauge invariance implements the closure condition $\sum_j x_j \!=\! 0$. 
We now propose a natural modification of the GFT (\ref{extaction}) in terms of a constraint operator acting on the field by implementing 
the simplicity constraint of its bivector variables, for any positive value of the Immirzi parameter $\gamma$, allowing to 
reconstruct a discrete tetrad for the tetrahedron. By construction, the Feynman expansion will generate simplicial path integrals for a constrained BF theory of Holst-Plebanski type. 

We start by recalling the discrete form of the (linear) simplicity constraints for a classical bivector geometry. 

\subsection{Discrete simplicity constraints}

In the absence of Immirzi parameter, the simplicity constraints  state that the hodge dual $\ast x_j^{IJ}$ are the area bivectors of a geometric (metric) tetrahedron:  
this is the discrete equivalent of $B \!=\! \ast e\wedge e$. 
Following \cite{sergei, FK},  these constraints are implemented by  requiring that the four $\ast x_j^{IJ}$ lie in the same hypersurface normal to a given unit vector $k^I$ in $\R^4$, namely $\ast x_j^{IJ}k_J \!=\! 0$ for all $j\!=\!1...4$. Using the selfdual/anti-selfdual decomposition of the algebra $\so(4)$, this can be expressed as: 
\beq \label{simpnogamma}
\forall j\in \{1...4\}, \quad \exists k \in \SU(2), \quad  kx^\minus_j k^{\minus 1} + x^\plus_j = 0
\eeq
The  variable $k\in \SU(2)\!\sim\!\cS^3$ is then the $\SU(2)$ representation of the unit vector $k^I$ normal to the tetrahedron\footnote{Let $\bar{k}\!:=\!(\bar{k}^\minus, \bar{k}^\plus)$ be the $\SO(4)$ rotation mapping the vector $N^I\!=\!(1,0,0,0)$ to $k^I$, then $k\!=\!\bar{k}^\plus \bar{k}^{\minus 1}$.}. 
If ten bivectors labeling the faces of a 4-simplex satisfy simplicity and closure constraint for each tetrahedron, then they define a geometric 4-simplex  
(for non-degenerate configurations). Furthermore, if the (constrained) bivectors associated to a given tetrahedron are also correctly identified across the two 4-simplices sharing it, then the reconstruction of a discrete tetrad can be carried out for the whole simplicial complex, again modulo degenerate configurations.

The inclusion of the Immirzi parameter can be performed easily also at the discrete level \cite{FK}. 
We have mentioned that the Immirzi parameter is introduced in the continuum action (\ref{pleb}) by a change of variables $B\!\to\! \bar{B} \!=\! B+ \frac{1}{\gamma}\!\ast\! B$ in the BF term. The action in the new variables looks again like a constrained BF theory, but where the constraints $\cC(B(\bar{B}))$ are now imposed on the following linear combination field $\bar{B}$:
\beq  \label{change}
B(\bar{B}) \!=\!\frac{\gamma}{1- \gamma^2} (\ast \bar{B} - \gamma \bar{B})
\eeq 
In the discrete setting, the simplicity condition with Immirzi parameter is thus obtained from (\ref{simpnogamma}) by replacing  
by replacing $x_j$ by the linear combinations $\ast x^{IJ}_j - \gamma x^{IJ}_j$. In terms of the selfdual/anti-selfdual decomposition of the bivectors, it reads: 
\beq \label{simp}
\forall j\in \{1...4\}, \quad \exists k \in \SU(2), \quad  kx^\minus_j k^{\minus 1} + \beta x^\plus_j = 0
\eeq
where the parameter $\beta$ is related to the Immirzi parameter as:
\beq \beta\!=\!\frac{\gamma-1}{\gamma+1}\eeq 
Note that the relation (\ref{simp}) is invariant under simultaneous sign flip $\gamma \!\to\! -\gamma$ of the Immirzi parameter and exchange 
$x^\plus_j \leftrightarrow x^\minus_j$ of the self dual and anti-self dual part of the bivectors. 
It is also well-defined for $\gamma\!=\!1$, although then the change of variables (\ref{change})  is singular and the geometrical interpretation is lost. 
In the following, we restrict to $\gamma \!\in\! [0, \infty]$, so that  the parameter $\beta$ takes its value in $[-1, 1]$.

\subsection{Constraint operator and  non-commutative tetrahedra}

Back to GFT, we now need  to encode  the simplicity condition (\ref{simp}) of the bivector variables $x_j$ as a constraint on the  field $\vphihat_k$,  the idea being of course to identify the normal to the tetrahedron to the 
$\SU(2)$ variable $k$  of the field. 
The natural way to do so why  taking into account the non-commutativity of the fields is to use non-commutative delta functions, defined by their plane wave expansion (\ref{gaugeproj}). These delta functions act as distributions for the star product: $\delta\star \phi(a) \!=\! \phi(0) \delta(a)$, so using these to constrain the field will effectively amount to  
constrain the measure on the bivectors. 

We thus introduce the following function of $x\!=\!(x^\minus, x^\plus) \in \so(4)$:
\beq \label{simpfunct}
S_k^\beta(x) := \delta_{\minus kx^\minus k^\inv}(\beta x^\plus) = \int_{\SU(2)} \!\!\extd u \, \e^{i \mbox{\footnotesize tr} [k^\inv u k x^\minus]} \e^{i \beta  \mbox{\footnotesize tr} [u x^\plus]}
\eeq 
where $\delta_{\minus a}(b)\!:=\!\delta(a+b)$ and $\delta$ is the $\su(2)$ non-commutative delta function.  
Our geometrical GFT models will be defined by constraining the field $\vphihat_k(x_j)$ in the action (\ref{actionbiv}), by means of an operator 
$\widehat{S}^\beta$ acting on it  by $\star$-multiplication by the product
$S_k^\beta(x_1)...S_k^\beta(x_4)$ of four simplicity functions:
\beq \label{simplicity}
(\widehat{S}^\beta \acts \vphihat)_k(x_1, \cdots, x_4) = \prod_{j=1}^4 S_k^\beta(x_j) \star \vphihat_k (x_1,\dots x_4) 
\eeq
We give below the explicit expression of the star product  (\ref{simplicity}) in terms of group and tensor Fourier components.
But first, let us show that the action of this operator is well-defined. We will also see that it commutes with the $\SO(4)$ gauge transformations (\ref{extgaugebiv}). 

To be able to take the $\star$-product of the $S^\beta_k$ with the field, we need 
the function (\ref{simpfunct}) to be in the image of the group Fourier transform. To see why this is indeed the case, notice that, because $|\beta|\leq1$, there exists 
$u_\beta\!\in\! \SU(2)$ such that $\beta \mbox{tr}[au] \!=\!  \mbox{tr}[a u_{\beta}]$ for all $a\in \su(2)$. 
Indeed, if $u\!=\!\e^{\theta n^j \tau_j}$  is parametrized  by the angle $\theta \in [0,\pi]$ and the unit $\R^3$-vector $\vec{n}$, where the $\tau_j$ are i times the Pauli matrices, we define $u_\beta\!=\! e^{\theta_{\beta} n^j_{\beta} \tau_j}$, where the parameters $\theta_\beta$ and $\vec{n}_\beta$ are
\beq \label{ubeta}
\sin\theta_\beta \!=\! |\beta| \sin \theta, \,\, \mbox{\footnotesize sign}(\cos\theta_\beta) = \mbox{\footnotesize sign}(\cos\theta); \quad \vec{n}_{\beta} = \mbox{\footnotesize sign}(\beta) \vec{n}
\eeq 
The simplicity function (\ref{simpfunct}) can thus be written as a superposition of plane waves $E_g(x) \!=\! \e^{i \Tr gx}$:
\beq \label{simpexp}
\delta_{\minus  kx^\minus k^\inv}(\beta x^\plus) = \int_{\SU(2)} \!\!\extd u \, E_{\mathbf{u}^k_{\beta}}(x)
\eeq
where we introduced $\mathbf{u}^k_{\beta} \!=\! (k^{-1} u k, u_\beta) \!\in\! \SU(2) \!\times\! \SU(2)$. 
Therefore it belongs to the image of the Fourier transform, and its star-product with the field  is well-defined.  

The operator $\widehat{S}$ is not a projector for generic values of $\beta$, 
unless $\beta \!=\! 0, 1$ (which corresponds to $\gamma\!=\!1, \infty$).
Indeed, because of the nonlinearity of scaling by $\beta$ in the definition (\ref{ubeta}), 
we have that $(uv)_\beta \not\!=\! u_\beta v_\beta$, and thus $S^\beta_k \star S^\beta_k \not\!=\!S^\beta_k$. 
Remarkably, however,  the action of $\widehat{S}$ is well-defined on gauge invariant fields, as it commutes with the gauge transformations (\ref{extgauge}):
\beq
\widehat{S}^\beta\acts [E_h \cdots E_h \star \vphihat_{h^\inv\acts k}] = E_h \cdots E_h \star (\widehat{S}^\beta\acts\vphihat)_{h^\inv \acts k}
\eeq
thanks to the commutation relations between plane waves and simplicity functions:
  \beq
E_{h}  \star S^\beta_k = S^\beta_{h \rhd k} \star E_{h} 
\eeq
Geometrically, these relations express the fact that rotating a  bi-vector which is simple with respect to a normal $k$ gives  a bi-vector which is simple with respect to the rotated normal $h \acts k:=h^{\plus} k (h^\minus)^{\minus 1}$.  This is the advantage of the extended GFT formalism, where the normals are explicit variables of the field:  the linear simplicity constraints on the bivectors can be imposed a covariant way. 
This is not the case in the  standard formulation of the Barrett-Crane model, nor on the EPRL-FK model, where simplicity and gauge invariance are implemented by means of two non-commuting projectors.

Let us now examine the dual action of $\widehat{S}$ on the original group fields $\vphi_k(g_j)$.  By using the plane wave expansion (\ref{simpexp}) of the simplicity functions 
and the definition of the star product, we obtain: 
\beq \label{simplicitygroup}
(\widehat{S}^\beta\acts\vphi)_k(g_1, \cdots\!, g_4) = \int_{\SU(2)^4} \!\![\extd  u_j]^4 \, \vphi_k(\mathbf{u}^k_{1\beta} g_1 , \cdots \! ,\mathbf{u}^k_{4\beta} g_4)
\eeq
where $\mathbf{u}^k_{j\beta} \!=\! (k^{\minus 1} u_j k, u_{j\beta}) \!\in\! \SU(2) \!\times\! \SU(2)$ and 
$u_{j\beta}$ is defined as in (\ref{ubeta}).  For the particular value $\beta\!=\!1$ reached in the limit $\gamma \!\to\! \infty$, $\widehat{S}^1$ 
reduces to the projector onto fields on four copies of the homogeneous space $\SO(4)/\SO(3)_k$.
Using the invariance (\ref{extgauge}), one can gauge fix the normal to the value $k=1$ (time gauge).  
On such gauge fixed fields, the simplicity operator coincides with the projector  defining the standard GFT formulation of the Barrett-Crane model \cite{BC, danielearistideBC}. 
The difference here, however, is that
the gauge fixed extended fields are obviously not gauge invariant under the full $\SO(4)$ -- but only under the diagonal $\SU(2)$ subgroup. 

Upon Peter-Weyl decomposition of the constrained  field $(\widehat{S}^\beta\acts\vphi)_k$, 
a set of basis functions is given by the action of the $\widehat{S}$ on the functions (\ref{spinbasis}):
\beq \label{simplicityspin}
\widehat{S}^\beta \acts \Psi^{(J_i, k_i, j)}_{m^\minus_i, m^\plus_i}(g_i;k)= 
\left(\prod_{i=1}^4 D^{j_i^\minus}_{n^\minus_i m^\minus_i}(g_i^\minus) 
D^{j_i^\plus}_{n_i^\plus m_i^\plus}(g_i^\plus) F^{j_i^\minus j^\plus_i k_i}_{n^\minus_i n_i^\plus p_i}(k) \right)  (\iota_j)^{k_i}_{p_i}
\eeq
where repeated lower indices are  summed over. This expression is obtained from (\ref{spinbasis}) by replacing the k-dependent  coefficients 
$\widetilde{C}^{j^\minus j^\plus k}_{m^\minus m^\plus p}(k)\!=\!C^{j^\minus j^\plus k}_{m m^\plus p} D^{j^\minus}_{m m^\minus}(k)$ by new ones given by: 
\beq \label{simplicitycoefficients}
F^{j_i^\minus j^\plus_i k_i}_{n^\minus_i n_i^\plus p_i}(k)  = 
\int_{\SU(2)} \! \!\extd u \, D^{j^\minus_i}_{m^\minus_i n^\minus_i}(k^\inv u k) D^{j^\plus_i}_{m^\plus_i n^\plus_i}(u_\beta) \,
\widetilde{C}^{j_i^\minus j^\plus_i k_i}_{m^\minus_i m_i^\plus p_i}(k) 
\eeq
with $u_\beta$ given as in (\ref{ubeta}). 
Just as in (\ref{spinbasis}), these coefficients intertwine the action of stabilizer subgroup $\SO(3)_k$ in the representation $j^\minus_i \otimes j^\plus_i$ 
and the action of $\SO(3)$ in the representation $k_i$.  Namely, given $\mathbf{u}_k \!=\! (k^{\minus 1} u k, u)\!\in\! \SO(3)_k$, we have:
\beq \label{intF}
F^{j_i^\minus j^\plus_i k_i}_{m^\minus_i m_i^\plus p_i}(k) D^{j^\minus_i}_{m^\minus_i n^\minus_i}(\mathbf{u}_k^\minus) D^{j^\plus_i}_{m^\plus n^\plus_i}(\mathbf{u}_k^\plus)
= F^{j_i^\minus j^\plus_i k_i}_{n^\minus_i n_i^\plus q_i}(k) D^{k_{i}}_{q_i p_i}(u).
\eeq
 Here, they also contain all the information about the simplicity constraints and the specific form of the operator that implements them. In particular the integral of the two Wigner matrices encodes a relation between the spins $(j^\minus_i, j^\plus_i)$, which depends on the Immirzi parameter; for example $j^\minus \!=\! j^\plus$ when $\beta\!\in\!\{\minus 1, 1\}$, namely when $\gamma\in\{0,\infty\}$. However for generic values of $\beta$, 
it does not enforce the spin relations $j^\minus\!=\!|\beta| j^\plus$ characteristic of the EPRL-FK models, an analogue of which we expect to recover only in the asymptotic regime. In particular, they do not impose any rationality condition on the Immirzi parameter $\gamma$. 
A detailed study of the properties of these coefficients, in particular their asymptotic behaviour for large spins, is left for future work.

\

To sum up this section, we have defined an operator $\widehat{S}^\beta$ acting on the field by imposing the linear simplicity condition (\ref{simp}), 
for any positive value of the Immirzi parameter, on its four bivector variables. Note that, because of gauge invariance,  the closure constraint holds after integration over the normal: let $\widehat{\psi}\!:=\! \int\extd k \widehat{S}^\beta \acts \vphihat_k$, then $\widehat{\psi} \!=\! \delta(x_1+..+ x_4) \star \widehat{\psi}$ where $\delta$ is the non-commutative delta function $\delta\star \phi(x)\!=\!\phi(0)\delta(x)$ defined in (\ref{gaugeproj}).

In relation to a canonical theory, on cane link the definition of the constrained GFT field to a quantization of a tetrahedron characterized by its four constrained bivectors: 
in particular, one can check (using the Fourier duality with group fields)  that the generators $J^{IJ}_j$ of the gauge group act by $\star$-multiplication by the coordinate functions $\hat{x}^{IJ}_j(x_j) \!=\! x^{IJ}_j$.  
 The quantization procedure consists of first quantizing classical configurations $\{x_j\}\!\in\! \so(4)$, $k\!\in\!\SU(2)\!\sim\! \cS^3$ of bivectors and normal and then to impose geometricity (simplicity)
constraints at the quantum level.  The Hilbert space is the tensor product of $L^2(\SU(2))$ with
\[
\bigotimes_{i=1}^4 L^2_{\star}(\R^6) 
\]
where the $L^2_{\star}$ spaces, which also appear as state spaces in the flux representation of loop quantum gravity \cite{flux}, 
are spaces of functions on $\so(4)\!\sim\! \R^6$ endowed with the scalar product $\int \extd^6x (\bar{f}\star g)(x)$
where $\bar{f}(x)\!=\!f(-x)$. The algebra structure encoded in the star product, which deforms the usual point-wise product, 
stems directly by Fourier transform from the algebra of group functions. It makes explicit the non-commutativity of the geometry inherent to spin foams and group field theories.  This procedure is manifestly dual to geometric quantization: 
the advantage here is that the classical variables characterizing the geometry remain explicit, as arguments of the fields. 
Geometricity conditions are then implemented by using two commuting operators: the simplicity operator $\widehat{S}^\beta$ and the gauge projector $\widehat{C}$
defined by (\ref{gaugeproj}), leading to  the unambiguous definition of a `geometricity operator' 
$\widehat{G} \!=\!\widehat{S}^\beta\widehat{C} \!=\!\widehat{C} \widehat{S}^\beta$. 
All constraints are  imposed by means of non-commutative delta-functions, acting as Dirac distributions for the star product, so that
it effectively amounts to constrain the measures $\extd^6 x_j$ on the classical field variables. 

The algebra structure and scalar product allow to build up more involved polyhedra obtained by glueing tetrahedra along a common triangle. 
Thus, glueing five geometric tetrahedra  $\widehat{\Psi}\!:=\! \int\extd k \widehat{G}\acts \vphihat_k$ along common triangles as 
\beq
\widehat{\Psi}_{1234}\star \widehat{\Psi}_{\minus 4567}\star \widehat{\Psi}_{\minus 7\minus 3 8 9}\star \widehat{\Psi}_{\minus 9\minus 6\minus 2 10}
\star \widehat{\Psi}_{\minus 10\minus 8\minus5 \minus 1},
\eeq
where $\widehat{\Psi}_{k\pm 1 \pm 2\pm 3\pm4}$ is a shorthand notation for $\widehat{\Psi}_{k}(\pm x_1, \cdots, \pm x_4)$ and the star product pairs repeated lower indices, gives a straightforward  ansatz for a `quantum 4-simplex'. This is precisely the interaction polynomial of our geometrical GFT, which we define now.

\subsection{GFT for gravity with no Immirzi parameter}

We first focus on the particular case $\beta\!=\!1$. It corresponds to $\gamma\!\to\!\infty$, namely the case of gravity with no Immirzi parameter. 
In this case, the operator $\widehat{S}^1$ reduces to  the projector onto the homogeneous space $\SO(4)/\SO(3)_k$.
The corresponding geometrical GFT has been introduced and studied in \cite{danielearistideBC}; we recall the basics here, before treating the general case. 

The model is defined by constraining the field  $\vphihat_k$ in the action of the extended Ooguri model. 
Since $\widehat{S}^1$ is a projector, it can be inserted in the propagator, in the vertex, or in both -- without affecting the amplitudes. In particular, these choices all lead to a unique form of edge amplitudes in the spin foam model. 
Here we set $\widehat{\Psi}\!:=\! \int\extd k \widehat{S}^1\acts \vphihat_k$ and define the action: 
\beqa \label{actionBC}
S \!\!&=&\!\!\frac{1}{2} \int [\extd^6 x_i]^4 \,\extd k  \, \widehat{\varphi}_{k1234} \star \widehat{\varphi}_{k \minus 1 \minus 2 \minus 3\minus 4} \nonumber  \\
&& \hspace{1cm}  + \frac{\lambda}{5!}  \int [\extd^6 x_i]^{10}   \,
\widehat{\Psi}_{1234} \star \widehat{\Psi}_{\minus 4567}\star  \widehat{\Psi}_{\minus 7\minus 389}\star  \widehat{\Psi}_{\minus 9\minus 6\minus 2\,10} \star \widehat{\Psi}_{\minus 10\,\minus 8\minus 5\minus 1} 
\eeqa
where  the star product pairs repeated indices. 

We now calculate the Feynman amplitudes of this theory.  
We will obtain two dual representations of these, in terms of  simplicial path integrals on one hand, and spin foam models on the other.


\subsubsection{Simplicial path integral representation of the amplitudes} 

The Feynman amplitudes of this theory are calculated with the same propagator as in (\ref{propext}) and  the vertex: 
\beq
V(x_i^\ell; k_{\ell}) = \int  [\extd h_{\ell}]^5 \, \prod_{i=1}^{10} ( \delta_{\minus x^{\ell}_i} \star 
S_{k_\ell} \star\, E_{h_\ell h^{\minus 1}_{\ell'}} \star S_{k'_\ell})(x^{\ell'}_i)
\eeq
where $i$ labels the oriented strands (triangles) and $\ell$ the half lines (tetrahedra) of the vertex graph, and $S_k \!=\! \delta_{\minus kx^\minus k^\inv}(x^\plus)$ is the simplicity function  for $\beta\!=\!1$. The calculation is analogous to the unconstrained case. 
For a given closed graph dual to a simplicial complex $\Delta$,  this results in integrals over holonomies $h_{\tau\sigma}\!\in \SO(4)$, tetrahedra normals $k_{\tau}\!\in \SU(2)$, 
and bivectors $x_t^\tau, x_t^\sigma$ on $t$ seen in different frames associated to the tetrahedra and 4-simplices sharing $t$. 
After integration over all variables but one per triangle $x_t\!:=\!x_t^{\tau_0}$ associated to a reference tetrahedron $\tau_0(t)$, the amplitude reads \cite{danielearistideBC}: 
\beq \label{amplitudeBC}
\cI_{\Delta}  = \int  [\extd h_{\tau\sigma}][\extd k_{\tau}] [\extd^6 x_t] \left[\prod_t \bigstar_{j=0}^{N_t}  \, 
S_{h_{0j} \acts k_j}(x_t)\right]  \star e^{i \sum_t \Tr \, x_t H_t}
\eeq
The notations are that of Sec.\ref{bivrep}: $H_t\!=\!h_{\tau_0\tau_1} \cdots h_{\tau_{N_t}\tau_0}$ is the holonomy along the (oriented) loop of $N_t+1$ tetrahedra sharing $t$, labelled by the integer $j$;  $h_{0j}\!=\!h_{\tau_0\tau_1}\cdots h_{\tau_{j-1}\tau_j}$ is the holonomy from the reference tetrahedron to the $j$-th tetrahedron sharing $t$. 
$k_j\!:=\!k_{\tau_j}$ is the normal of the j-th tetrahedron around $t$. The function $S_{h_{0j} \acts k_j}(x_t)$ imposes on $x_t$ the linear simplicity condition with respect
to the rotated normal $h_{0j} \acts k_j\!:=\!h^\plus_{0j} k_j (h^\minus_{0j})^{\minus 1}$, namely the pull back of $k_j$ in the frame of the reference tetrahedron $\tau_0(t)$. 
It amounts to imposing the linear simplicity of the pushed forward bivector $h_{0j}^{\minus 1} x_t h_{0j}$ with respect to $k_j$. 
The integrand results from taking, for each $t$, the  alternate star product of $N_t$ plane waves and $N_t$ simplicity functions: 
\beq
S_{k_0} \star E_{h_{\tau_0 \tau_1}} \cdots \star S_{k_{N_t}}\star E_{h_{\tau_{N_t} \tau_0}}(x_t) = 
\left[\bigstar_{j=0}^{N_t}  S_{h_{0j} \acts k_j}\right]\star  e^{i\Tr \, x_t H_t}
\eeq
where we used the commutation relation $S_k\star E_h \!=\! S_{h\acts k} \star E_h$ to regroup all the plane waves on the right of the expression. 

The Feynman amplitudes of the GFT (\ref{actionbiv}) thus take the form of (non-commutative) simplicial path integrals for a constraint BF theory of Plebanski type. 
The constraints are non-commutative delta function modifying the measures $\extd^6 x_t$ on the bivectors, imposing the simplicity of each $x_t$ with respect to the normals of all the tetrahedra sharing $t$.

Note that by construction, the integrand of \ref{amplitudeBC} is invariant under $\SO(4)$ rotations $\{g_{\tau}, g_{\sigma}\}$ of all local frames:
\beq \label{gauge}
h_{\tau\sigma} \mapsto g_\tau h_{\tau\sigma}g_{\sigma},\qquad
k_\tau \mapsto g^\plus_\tau k_{\tau} (g_{\tau}^{\minus})^{\minus 1}\qquad
x_t \mapsto g^{\minus 1}_{\tau(t)} x_t g_{\tau(t)}
\eeq
where $\tau(t)$ is the reference tetrahedron of the triangle $t$.  This includes the gauge invariance of the discrete $BF$ action. 
The choice $g_{\tau} \!=\! (g^\minus_\tau, g^\plus_{\tau}) \!:=\! (k^{\minus 1}_\tau,1)$ 
leads to the `time gauge' $k_{\tau}\!=\!1$, which shows that the integral over the normals $k_{\tau}$ drop out of the amplitude. 
Of course in the case of open graphs,  dual to simplicial complexes with boundary, the amplitude still has an explicit dependence on the normals of  the boundary tetrahedra.

It is interesting to distinguish two types of constraints on the bivector $x_t$ of a given triangle.  
The $j\!=\!0$ contribution $S_{k_0}(x_t)\!=\!\delta_{\minus k_0 x^\minus_t k_0^\inv}(x^\plus_t)$ in (\ref{amplitudeBC}) 
imposes the linear simplicity of each bivector $x_t$ with respect to the normal to the reference tetrahedron $\tau_0(t)$. 
The remaining part, for a given set of bivectors, can be viewed as constraints on the holonomies  modifying  the measures $\extd h_{\tau \sigma}$ on the discrete connections. The effective measure
\beq \label{holmeasure}
\cD^{x_t,k_{\tau}}[h_{\tau\sigma}]= [\extd h_{\tau \sigma }] \prod_t \vec{\bigstar}_{j=1}^{N_t}  \delta_{\minus h_{0j} \acts  k_j x^\minus_t (h_{0j} \acts  k_j)^\inv}(x^\plus_t)
\eeq
transforms covariantly under the gauge transformations (\ref{gauge}). Using the cyclic invariance of the $\star$-product under integration, the amplitude (\ref{amplitudeBC}) can be written in terms of this measure as: 
 \beq 
\cI_{\Delta} \!=\! \int [\extd^6x_t][\extd k_{\tau}] \cD^{x_t, k_\tau}[h_{\tau\sigma}] \star \prod_t \left[e^{i  \Tr \, x_t H_t} \star 
\delta_{\minus k_{\tau_o(t)} x^\minus_t k_{\tau_o(t)}^\inv}(x^\plus_t)\right]
\eeq
where the star product pairs the variables $x_t$. 
Note that the appearance of a gauge covariant measure on the discrete connection is the result of using extended GFT fields, 
and thus of requiring a covariant imposition of the simplicity constraints. 
The need of a  generalization of the closure constraint to achieve this has been noted on several occasions in the literature \cite{sergei, danielesteffen} (see also \cite{Plebcanonical}).

\subsubsection{Spin foam representation of the amplitudes}

We have derived the GF amplitudes  starting from the bivector formulation of the group field theory, where the simplicial geometry is implicit. 
By construction, they take the form of simplicial path integrals. 
It should be clear, however, that the dual connection and spin formulations of the  same GFT will dual expressions of the same amplitudes in terms of a lattice gauge theory and spin foam amplitudes. 

The spin foam representation of the amplitudes (\ref{amplitudeBC}) can be also be computed directly by Plancherel decomposition of the group functions into irreducible representations and integration over group and Lie algebra elements. As proved in \cite{danielearistideBC}, it gives the  Barrett-Crane amplitudes:
\beq \label{amplitudemeas}
\cI_{\Delta} = \cI_{BC} :=  \sum_{\{j_t\}} \prod_t d^2_{j_t} \prod_\tau \frac{1}{\prod_{t \in \partial\tau} d_{j_t}}  \prod_{\sigma} \{10j\}_{\sigma} 
\eeq
The sum is over $\SU(2)$ spins $j_t$ labeled by triangles; $d_{j_t}\!:=\!2j_t+1$. The products are over all triangles $t$, tetrahedra $\tau$ and  4-simplices $\sigma$. 
The 4-simplex weight  $\{10j\}_{\sigma}$  is the Barrett-Crane $10j$-symbol \cite{BC}. 
This derivation singles out a specific edge amplitude (tetrahedral weight), which differs from the ones that appear in the literature with the exception of \cite{eteravalentin}. We emphasize again that, in our GFT construction, although the amplitudes of closed graphs reproduce the Barrett-Crane amplitudes, 
the requirement of a covariant imposition of the simplicity constraints imposed the use of extended boundary states which include tetrahedra normals, 
hence labelled by projected spin networks. 

This geometrical construction sheds an interesting new light on the Barrett-Crane model. 
In particular, the formula (\ref{amplitudeBC}) gives a new simplicial path integral formulation of its amplitudes, making the simplicial geometry of the model manifest.  
We refer to \cite{danielearistideBC} for a detailed analysis and discussion.  

\subsection{Including the Immirzi parameter: GFT for Holst-Plebanski gravity}

We now turn to the inclusion of the Immirzi parameter in the model. 
For generic values of $\beta\!=\!\frac{\gamma-1}{\gamma+1}$,  the operator $\widehat{S}^\beta$ is no longer a projector. 
Depending on whether it is inserted in the propagator, in the vertex, or in both the vertex and propagator, in a single of in multiple copies, 
will lead to a priori different spin foam amplitudes. These will however have the same vertex amplitude (4-simplex weight) and differ only 
in the edge and face amplitudes (weights associated to tetrahedra and triangles). 
Here, just as in the previous section, we constrain the field in the interaction of the extended Ooguri model. Setting 
$\widehat{\Psi}^\beta\!:=\! \int\extd k \widehat{S}^\beta\acts \vphihat_k$, we thus consider: 
\beqa \label{actionbeta}
S \!\!&=&\!\!\frac{1}{2} \int [\extd^6 x_i]^4 \,\extd k  \, \widehat{\varphi}_{k1234} \star \widehat{\varphi}_{k \minus 1 \minus 2 \minus 3\minus 4} \nonumber  \\
&& \hspace{1cm}  + \frac{\lambda}{5!}  \int [\extd^6 x_i]^{10}   \,
\widehat{\Psi}_{1234} \star \widehat{\Psi}_{\minus 4567}\star  \widehat{\Psi}_{\minus 7\minus 389}\star  \widehat{\Psi}_{\minus 9\minus 6\minus 2\,10} \star \widehat{\Psi}_{\minus 10\,\minus 8\minus 5\minus 1} 
\eeqa
where  the star product pairs repeated indices. 

\subsubsection{Simplicial path integral representation of the amplitudes} 

The derivation of the Feynman amplitudes is analogous to the case $\beta\!=\!1$. They are calculated 
with the same propagator as in (\ref{propext}) and  the vertex: 
\beq
V^{\beta}(x_i^\ell; k_{\ell}) = \int  [\extd h_{\ell}]^5 \, \prod_{i=1}^{10} ( \delta_{\minus x^{\ell}_i} \star 
S^\beta_{k_\ell} \star\, E_{h_\ell h^{\minus 1}_{\ell'}} \star S^\beta_{k'_\ell})(x^{\ell'}_i)
\eeq
where $i$ labels the oriented strands (triangles) and $\ell$ the half lines (tetrahedra) of the vertex graph, and $S^\beta_k \!=\! \delta_{\minus kx^\minus k^\inv}(\beta x^\plus)$ is the simplicity function.  
Using the same notations as in (\ref{amplitudeBC}), we obtain, for the amplitude of a closed graph: 
\beq \label{amplitudebeta}
\cI^{\beta}_{\Delta}   = \int  [\extd h_{\tau\sigma}][\extd k_{\tau}] [\extd^6 x_t] \left[\prod_t \bigstar_{j=0}^{N_t}  \, 
S^{\beta \star 2}  _{h_{0j} \acts k_j}\right]  \star e^{i \sum_t \Tr \, x_t H_t}
\eeq
where $S^{\beta \star 2}$ denotes the squared function $S^{\beta} \star S^\beta$. 
For each triangle $t$, the constraints impose, by means of non-commutative delta functions,  
the linear simplicity condition of $x_t$ with respect to the normals of all the tetrahedra $\{\tau_j\}_{j=0...N_t}$ sharing $t$.  
The square stems from the fact both  4-simplices sharing the tetrahedron $\tau_j$ contributes to a factor $S^{\beta}  _{h_{0j} \acts k_j}(x_t)$. 
The  Feynman amplitudes of this theory thus take the form of simplicial path integrals for a constrained BF theory of Holst-Plebanski type with Immirzi parameter $\gamma$, with linear simplicity constraints \cite{danielesteffen}. 

Just as for the case with no Immirzi parameter, the integrand of (\ref{amplitudebeta}) are invariant under the gauge transformations (\ref{gauge});  
in the case of closed graphs, the integration over the normals $k_{\tau} \!\in\! \SU(2)$ drops from the amplitude. 
Making explicit the form of the simplicity functions, we thus obtain: 
\beq \label{explPI}
\cI^{\beta}_{\Delta} = \int  [\extd h_{\tau\sigma}][\extd^6 x_t] \left[\prod_t \bigstar_{j=0}^{N_t} 
\delta^{\star 2}_{\minus \bar{h}_{0j} x^\minus_t  \bar{h}_{0j} ^\inv}(\beta x^\plus_t)\right]  \star e^{i \sum_t \Tr \, x_t H_t}
\eeq
where we wrote $f^{\star 2}$ for the squared function $f\star f$, and $\bar{h}_{0j}\!=\!h_{0j}^\plus (h^\minus_{0j})^{\minus 1}$. 
Finally, by splitting the constraints into a part (the $j\!=\!0$ contributions) that is independent of the holonomies and a part playing the role of constrains on the holonomies, the amplitude $\cI^{\beta}_{\Delta}$ can be put under a form analogous to (\ref{amplitudemeas}), in terms of a covariant measure on the space of discrete connections. 

\subsubsection{Spin foam representation of the amplitudes}

The spin foam representation of the amplitudes can be obtained either directly from (\ref{amplitudebeta})  
by inverse Fourier transform and Peter-Weyl decomposition of the group functions, or from the Feynman rules of the spin representation of the generating GFT. 
In this section we give the explicit form of the resulting spin foam amplitudes in terms of 15j-symbols and so-called fusion coefficients, which will allow a direct comparison with the existing models \cite{EPRandL, FK, FKpathintegral}. Their derivation from the GFT is straightforward:  we only sketch it here.  

In the spin foam representation, the amplitudes read:
\beq \label{spinfoam}
\cI^{\beta}_{\Delta} = \sum_{j^\minus_t, j^\plus_t, \lambda_{t \tau}, \imath_\tau } \prod_{t} d_{j_t^\minus} d_{j_t^\plus} \prod_{(t \tau)} d_{\lambda_{t\tau}} 
\prod_{\sigma}  A^\beta_\sigma(j^\pm_t, \lambda_t,  i_{\tau}; k_{\tau})
\eeq
where  the 4-simplex weight (vertex amplitude) is given by: 
\beq
A^\beta_\sigma(j^\pm_t, \lambda_{t\tau},  i_{\tau}; k_{\tau})
= \sum_{\imath_{\tau \sigma}^\minus, \imath_{\tau \sigma}^\plus} \{15j\}^\minus_{\sigma}\{15j\}^\plus_{\sigma}\prod_{\tau\subset \sigma} 
d_{\imath_{\tau\sigma}^\minus}  d_{\imath_{\tau\sigma}^\plus} f^{\imath_\tau}_{\imath^\minus_{\tau\sigma}, \imath^\plus_{\tau\sigma}}(j^\pm_t, \lambda_{t\tau}; k_{\tau})
\eeq
The notations are as follows. $t, \tau, \sigma$ denote the triangles, tetrahedra and 4-simplices of the simplicial complex $\Delta$. 
The sums are over $\SO(3)$ representation $j, \lambda$ and four-valent $\SO(3)$ intertwiners $\imath$, all labelled by an integer spin, and $d_j\!=\!2j+1$.
This gives a pair of spins $(j^\minus_t, j^\plus_t)$ for each triangle,  
a spin $\lambda_{t \tau}$ for each couple $(t\tau)$ with $t\subset \tau$, a
a spin $\imath_{\tau\sigma}$ for each tetrahedron and a pair of spins $(\imath^\minus_{\tau}, \imath^\plus_{\tau})$ for each couple $(\tau\sigma)$ with $\tau\subset \sigma$. We set $d_j\!=\!2j+1$. The variables $k_{\tau}\!\in\! \SU(2)$ are the normals to the tetrahedra: 
as we have seen, the dependence upon the normals for the bulk tetrahedra (internal links of the GFT graph) drop, 
hence we haven made the integrals over these explicit. The above amplitude may as well be evaluated in the time gauge $k_{\tau}\!=\!1$ for all bulk tetrahedra, 
though it then makes less transparent the nature the boundary states, here labelled by projected spin networks. 

The amplitude $A^\beta_\sigma$ is defined in terms of $\SU(2)$ Wigner symbols 
$\{15j\}^\pm_{\sigma}(j^\pm_t, \imath_{\tau\sigma}^\pm)$ and so-called fusion coefficients \cite{EPRandL} 
$f^{\imath_\tau}_{\imath^\minus_{\tau\sigma}, \imath^\plus_{\tau\sigma}}$. 
These coefficients define a map  from the space of $\SO(3)$ intertwiners between the representations $\lambda_{t_1\tau},..,\lambda_{t_4\tau}$ and the space of $\SO(4)$  intertwiners between the representations $(j_1^\minus, j_1^\plus),.., (j_4^\minus,j_4^\plus)$:
\beq
f|\imath_{\tau}\rangle \!=\! \sum_{\imath^\minus_{\tau\sigma}, \imath^\plus_{\tau\sigma}} f^{\imath_\tau}_{\imath^\minus_{\tau\sigma}, \imath^\plus_{\tau\sigma}} 
|\imath^\minus_{\tau\sigma} \otimes\imath^\plus_{\tau\sigma}\rangle
\eeq
While the form (\ref{spinfoam}) is quite general for a spin foam model defined a a constrained BF theory, the specificity of a model lies into the exact form of the fusion coefficients, which encode the way simplicity constraints are imposed.  For the new model presented here, they are given by:
\beqa
f^{\imath_\tau}_{\imath^\minus_{\tau\sigma}, \imath^\plus_{\tau\sigma}}(j^\pm_t, \lambda_{t\tau}; k_{\tau})
 &=& \langle \imath^{\minus}_{\tau\sigma} \otimes  \imath^{\plus}_{\tau\sigma}| \otimes_i F^{j_{t_i}^\minus j^\plus_{t_i} \lambda_{t_i \tau}}(k_\tau) | \imath_{\tau} \rangle
  \nonumber \\ 
  &=& (\imath^{\minus}_{\tau\sigma})_{m_i^\minus}  (\imath^{\plus}_{\tau\sigma})_{m_i^\plus} \left(\prod_i 
  F^{j_{t_i}^\minus j^\plus_{t_i} \lambda_{t_i \tau}}_{m^\minus_i m_i^\plus p_i}(k_{\tau})\right) (\imath_{\tau})_{p_i}
\eeqa  where repeated lower indices are summed over. 
The tensor $F^{j_{t_i}^\minus j^\plus_{t_i} \lambda_{t_i \tau}}(k_\tau)$ are the  ones defined in (\ref{simplicitycoefficients}). 
They provide an embedding of $\SO(3)$ structures into $\SO(4)$ ones. In particular, because of the intertwining property (\ref{intF}), 
they realize $\SO(3)$  as the stabilizer subgroup $\SO(3)_k\!\subset\!\SO(4)$ of the normal $k_{\tau}$ to the tetrahedron. 
They also depend on the Immirzi parameter and encode the simplicity constraints.

This form (\ref{spinfoam}) of the amplitudes follow from the GFT Feynman rules in the spin representation. 
In this representation, the bivectors $x_t$ are replaced by pairs of spins $J_t:=\!(j_t^\minus, j_t^\plus)$ and magnetic numbers labeling the strands of the graphs. 
A way to read these rules is then the following. For a given graph $\cG$ labelled by spins  $J_t$ and tetrahedron normals $k_{\tau}$,   
they attach to each 4-stranded line (tetrahedron) a propagator $P^{\beta, J_i}(k_{\tau})\in \mbox{End}(\bigotimes_{i=1}^4 J_{t_i})$
defined as  an endomorphism of the tensor product of the representations labeling its strands (triangles). 
The amplitudes are obtained by taking the trace (ie. index contractions) of all propagators, following the combinatorics of the graph and 
by summing over all spins (and normals) as:
\beq \label{pres}
\cI^{\beta}_{\Delta} = \sum_{\{J_t\}} \prod_{t} d_{j_t^\minus} d_{j_t^\plus} \,  \Tr_{\cG} \left[\bigotimes_{\tau} P^{\beta, J_i}(k_{\tau}) \right]
\eeq
where $d_j\!=\!2j+1$. The propagator decomposes as 
\beq
P^{\beta, J_i}(k_{\tau}) = P^{J_i}_{\SO(4)}  \widetilde{P}^{\beta, J_i}(k) P^{J_i}_{\SO(4)}
\eeq where 
$P^{J_i}_{\SO(4)}$ is the projector onto $\SO(4)$-invariant tensors  and $\widetilde{P}^{\beta, J_i}(k)$ is defined in terms of the tensors (\ref{simplicitycoefficients})  as:  
\beq  \label{prop}
\widetilde{P}^{\beta, J_i}_{m^\minus_i, m^\plus_i; n^\minus_i, n^\plus_i}(k)  = \prod_{i=1}^4 
\left( \sum_{\lambda_i} d_{\lambda_i}\overline{F}^{j_i^\minus j^\plus_i \lambda_i}_{m^\minus_i m_i^\plus p_i}(k) 
F^{j_i^\minus j^\plus_i \lambda_i}_{n^\minus_i n_i^\plus q_i}(k)\right) (P^{\lambda_i}_{\SO(3)})_{p_i q_i}
 \eeq
where repeated lower indices are summed over. $P^{k_i}_{\SO(3)}$ is the  projector onto $\SO(3)$ invariant tensors in $\otimes_{i} \lambda_i$. Note that its insertion is actually redundant in the definition of the propagator: indeed, because of the property (\ref{intF}), an $\SO(3)$ rotation in $\otimes_{i} \lambda_i$ is intertwined by $F$ with a $\SO(3)_k$ rotation in $\otimes_{i} J_i$, which can be reabsorbed into  $P^{J_i}_{\SO(4)}$.
This form however allows us to split the trace in (\ref{pres}) into a product of 4-simplex weights as in (\ref{spinfoam}). This is done by expanding the projectors $P^{J_i}_{\SO(4)}$ and $P^{k_i}_{\SO(3)}$ into  four valent intertwiners $(\imath^\minus, \imath^\plus)$ and $\imath$.

\

To close this section, we emphasize again that, as in the extended BF case, the boundary states of the model are by construction (constrained) projected spin networks \cite{projected}. 
Even in the presence of the Immirzi parameter, we see therefore that the boundary states of the amplitudes (or the GFT polynomial observables) are different from the states of standard LQG. The projected spin network structure is actually present also in the boundary states of all the new models \cite{EPRandL, FK}, even if in a less explicit way, and even if their apparent coincidence with LQG states in representation space (due to the specific form of the simplicity constraints imposed there) is more emphasized.

\subsection{Limiting cases}

The model presented in the previous sections corresponds to a candidate quantization of a simplicial version of the Plebanski-Holst formulation of 4d gravity, for generic values of the Immirzi  parameter $\gamma$.  We have already seen that the case $\beta\!=\!1$, which corresponds to $\gamma\!=\!\infty$, gives a variant of the Barrett-Crane model with a specific edge amplitude, where the boundary states are extended to include tetrahedra normals, hence are  labelled by projected spin networks. 
Thus, the formula (\ref{amplitudeBC}) not only gives a new simplicial path integral formulation of the BC model, 
but it also provides a natural deformation of that model which includes the Immirzi parameter. 
We now discuss briefly other limiting cases of this model: the `self-dual' case $\gamma\!=1\!$ and the `topological' case $\gamma \!=\! 0$.

For $\gamma\!=\!1$, as mentioned the change of variables \ref{change} becomes singular -- so the contact with the classical Holst theory is lost.  
Despite the lack of a clear geometric interpretation,  the constraint operator $S^{\beta}$ and the resulting model are well-defined for $\beta\!=\!0$; 
it acts on the field by $\vphihat_k$ by projection of its bivector variables onto the selfdual part of $\so(4)$. 
The constrained model reduces to the Ooguri model for topological $\SU(2)$ BF theory.

The case $\gamma \!=\! 0$ corresponds to the so-called topological sector of Holst gravity. This denomination comes from the  fact that the term of the classical Holst action that seemingly dominates in this limit is the one that vanishes on shell, due to the requirement of torsion free-ness of the connection. As a consequence, one would expect that the resulting spin foam model/path integral would define a trivial dynamics for any boundary state. It is not totally obvious, however, that the above reasoning goes through in the quantum theory as well. It could also be argued \cite{dariosimone} that the resulting quantum theory would rather correspond to a quantization of 2nd order, metric gravity with no torsion. The rough argument is that in a path integral for the Holst action, the limit $\gamma\rightarrow 0$ would force, analogously to a semi-classical limit, the same path integral to be dominated by solutions of the equations of motion coming from the \lq\lq topological term\rq\rq only, that is exactly the torsion free-ness condition. While these arguments are obviously not conclusive, they suggest not to dismiss the resulting model as un-interesting.
In our context, this corresponds to $\beta \!=\! -1$. 
As discussed also in \cite{danielearistideBC}, the constraint operator for $\beta \!=\! -1$:
a) projects onto simple $\SO(4)$  representation $J=(j,j)$, b) does not impose any restriction on the expansion of $(j,j)$ into $\SU(2)$ irreducible representations  
$k=0, \cdots 2j$, and acts on each component $(J, k)$ by multiplication by the phase $(-1)^{2j+k}$. 
In computing the amplitudes, the phase factors cancel each other. The resulting model,  distinct from the EPR amplitudes \cite{EPRandL}, is obtained from the $\SO(4)$ 
Ooguri model (\ref{ooguri}) by restriction the representations to simple ones $J_t\!=\!(j_t, j_t)$. It would be interesting to study what the geometric interpretation of such amplitudes may be.

The general GFT model for arbitrary $\beta$ thus encompasses and generalizes several distinct models, and interpolates between them. It is tempting to speculate (see also \cite{GFTrenorm, dariosimone}), that the model possesses a non-trivial renormalization group flow in parameter space $(G_N,\beta)$, where $G_N$ is the Newton's constant, which is hidden in our formulation as we use dimensionless quantities throughout. The natural candidates for fixed points would then be these special values for the $\beta$ parameter: $(-1, 0, 1)$, namely $\gamma\!=\!0, 1$ or $\infty$.  In particular, while the case $\beta \!=\! -1$ is distinguished only for being in some sense \lq extremal\rq, and for its peculiar classical analogue, the other two values can be seen as special already at the level of the very definition of the corresponding model. In fact, as we have seen, in these two cases, and only then, the simplicity operator defines a projector, and the quantum amplitudes  are insensitive to the specific choice of insertion of this operator in the GFT action.

\section{Simplex correlations and ultralocality}
\label{ultralocality}

The main advantage of the framework developed in this paper is that the variables encoding the (fuzzy) bivector simplicial geometry of GFT and spin foam models are explicit. 
In particular, with respect to other formalisms, it gives a more direct access to the way simplices are correlated in the model, namely, 
how the model relates the geometrical data of common subsimplices in the glueing of neighboring simplices.

It has in fact often been argued that, for example, the Barrett-Crane model, suffers from a default of correlations between neighboring simplices. 
This `ultralocality' feature has been one of the reasons to discard this model in favour of the new models. 
Since the Barrett-Crane amplitudes show up in our model for the value $\gamma\!=\!\infty$ of the Immirzi parameter, this feature can in principle be clarified and dwelled further. An indepth study of the ultralocality issue is beyond the scope of the present paper.  
In this section we however discuss how it manifests itself in our framework. 
 

The interaction and kinetic polynomials of our GFT model are written as  a simple star product of copies of the constrained field in which 
the bivector variables associated to the common triangles are {\sl strictly} identified (modulo an orientation flip). 
At the level of the Feynman amplitudes however, after expansion of the gauge invariance operator, 
the Feynman rules dictates the relation between bivectors $\{x_t^\tau, x_t^\sigma\}$ expressed in different frames related by holonomies $h_{\tau\sigma}$. 
For example in the  $BF$ model,  
two bivectors $x^\tau_t, x^{\tau'}_{t}$ on the same triangle $t$ but 
seen from different tetrahedra are related by a non-commutative delta function:
\beq \label{strand}
(\delta_{\minus x^\tau_t} \star E_{h_{\tau\tau'}} )(x^{\tau'}_t)
\eeq
where $h_{\tau\tau'}\!=\!h_{\tau\sigma}h_{\sigma\tau'}$ parallel transports the frame of one tetrahedron to the frame of the other. 
The structure of the star product gives a clear geometrical meaning to the algebraic expressions. 
In particular, the algebraic operation corresponding to the parallel transport of bivectors is the commutation with plane waves: $E_h \star f \! =\! f^h \star E_h$, where $f^h(x) \!=\! f(h^{\minus 1} x h)$. 

In the amplitudes of the constrained theory, the plane waves are supplemented with simplicity functions imposing the linear simplicity condition of the bivectors in each frame:
\beq \label{strandconstrained}
E_{h_{\tau\tau'}} \quad \longrightarrow \quad  S^\beta_{k_{\tau}}(x_t)  \star E_{h_{\tau\tau'}} \star S^\beta_{k_{\tau'}}(x_t)
\eeq
so that the commutation with the plane wave encodes the parallel transport of {\sl simple} bivectors. 
In the case $\beta\!=\!1$ (ie $\gamma \!=\! \infty$) corresponding to the Barrett-Crane amplitudes, the simplicity functions $S_{k}\!:=\! S^1_{k}$ satisfy  $S_{k}\star S_k\!=\!S_k$.  
The definition (\ref{simpfunct}) of these functions and the structure of the star product then lead to the identity:
\beq
S_{k_{\tau}}  \star E_{h_{\tau\tau'}} \star S_{k_{\tau'}} =  S_{k_{\tau}}  \star E_{\mathbf{u}_t^\tau h_{\tau\tau'} \mathbf{u}_t^{\tau'}} \star S_{k_{\tau'}}
\eeq
for all  $\bold{u}_t^{\tau}\!=\!(k_\tau^{\minus 1} u_t^\tau k_\tau, u_t^\tau)$ and $\bold{u}_t^{\tau'}\!=\!(k_{\tau'}^{\minus 1} u_t^{\tau'} k_{\tau'}, u_t^{\tau'})$ in the stabilizer subgroups $\SO(3)_{k_{\tau}}$ and $\SO(3)_{k_{\tau'}}$ of the normals. 
This is because $\bold{u}_t^{\tau}, \bold{u}_t^{\tau'}$ can be reabsorbed into the group elements labeling the plane wave expansion (\ref{simpfunct})  of $S_{k_\tau}$ and $S_{k_{\tau'}}$. This identity can be understood as a relaxation of the parallel transport condition, 
or a weakening of the bivector correlations: upon parallel transport  $h_{\tau \tau'}$, simple bivectors are identified only up to spatial rotations. 
Only seems to remain manifest the coupling of the bivector norms, i.e the area of the triangles.  
This is the manifestation of `ultralocality' in this geometrical setting, here due to  
the interplay between simplicity and parallel transport conditions induced by the non-commutativity of the star product. 

Note that the argument does not extend to general values of $\beta$ in an obvious way, for the same reason that makes the constraint operator fail be a projector 
$S_k\star S_k\!\not=\!S_k$. 
However it remains that upon commutation with \ref{strandconstrained}, a Lie algebra function gets conjugated not only by the holonomy $h_{\tau\tau'}$, but also by the Lagrangian multipliers of the simplicity functions. 

Arguing whether or not this feature is a serious problem from the point of view of quantum geometry is not our point here. 
Our point is to emphasize that it appears in our framework  as an unavoidable feature following a clear geometrical construction of a dynamical theory for non-commutative tetrahedra. 
In fact, as it has been anticipated in the literature, the above argument shows that it is manifestly inherent to the non-commutativity of the bivector geometry, here entirely encoded into the star product.  From the point of view of the canonical theory, where the tetrahedron states live in a tensor product of non-commutative spaces $L^2_\star (\R^6)$,  it is tied to the choice of quantization map. 
The presence of the star product thus encodes also quantum corrections, of which the above effect is a manifestation. 

This raises the question whether ultralocality survives in a semi-classical regime involving a commutative limit. The star product structure being dual to group composition, this limit corresponds to a linearization of the group.  
It can be formally defined by introducing a parameter $\epsilon$ in the coordinates of the group manifold and to parametrize  
$\SU(2)^\pm$ group elements $u\!=\!\e^{i\theta\vec{n}.\tau}$ for e.g by $\R^3$ vectors $\vec{p}_u \!=\! \frac{1}{\epsilon} \sin\theta \vec{n}$. 
The Fourier transform can be parametrized accordingly   \cite{PR3}; 
in the regime of small  $\epsilon$, the $\SO(3)$ (resp. $\SO(4)$) star product reduces to the usual pointwise product on functions of $\R^3$ (resp. $\R^6$). 
The commutative regime should correspond by duality to the large spin limit of spin foam models. 
However since it amounts to linearize the holonomies, it could also be viewed as an analogue of the continuum limit in lattice gauge theories. 
Whether or not a proper development  of the model around a commutative limit, at the level of the GFT  \cite{abeliangft} or its amplitudes, can be properly defined, and shown to tame the ultralocality feature, remains to be seen.


\section{Conclusions}
\label{concl}

In this paper, we exploited a dual formulation of group field theories in terms of non-commutative bivector variables,  
which provide a duality between spin foam models and simplicial path integrals for constrained BF theories, to 
derive a new model for 4d gravity with Immirzi parameter. 
All geometrical variables remain explicit in this construction, which consists of  inserting a constraint operator in a GFT for 4d $\SO(4)$ BF theory 
implementing the discrete simplicity constraints turning quantum simplicial BF theory into quantum simplicial gravity. 

Thanks to the framework chosen, we can keep the geometric content of
all variables and of imposed constraints manifest at all stages of the
construction. The resulting amplitudes for each simplicial complex,
generated in the Feynman expansion of the GFT, give a quantum simplicial version of the Holst-Plebanski formulation of gravity.
We formulated these amplitudes both as BF simplicial path integrals with explicit non-commutative $B$ variables and in terms of Wigner 15j-symbols and fusion coefficients. The new model differs from existing ones in the literature; it imposes a different restriction on representation labels for
quantum states. In particular, it does not lead any rationality condition for the Immirzi parameter.

In light of this geometrical framework, we suggested a possible new perspective
on the issue of the quantum correlations between neighboring
simplices, often argued to be a problematic feature, for example, in
the Barrett-Crane model.  In our formalism, in fact, the relaxation of parallel transport condition
is an unavoidable consequence of the very non-commutative nature of bivector variables and of the
simplicity constraints, and tied to their quantization. 
Moreover, our framework is best suited for studying the geometric interpretation and consequences of this relaxation, as well as the semi-classical limit of the amplitudes. 

While this problematic issue certainly needs to be investigated
further, we believe that the construction we performed suggests to
consider the resulting model seriously as a candidate model for
quantum gravity. In fact, its features appear all natural from the point of view of quantum geometry. Of course, this new model should now to be tested in
all its aspects to support further or refute its validity. 
Note that its explicit formulation as a path integral for constrained BF theory, in contrast with the other existing models, should facilitate the study of its relation with a path integral quantization of continuum Holst-Plebanski gravity \cite{Plebcanonical, sergei}. 

We also believe that the non-commutative formalism on which the
construction is based, and that has proven useful already in different
contexts and for different purposes, should itself be dwelled into in
depth, to unravel even more aspects of (simplicial) quantum geometry. We have in mind the issue of symmetries, in particular the simplicial analogue of diffeomorphisms symmetry. This has been studied in the BF context in \cite{diffeoGFT} and the analysis should now be extended to the 4d gravity model proposed here. Indeed, the bivector representation is the most suited one for defining such symmetries in a geometrically clear way.

Finally, the really crucial question is whether it leads to an effective continuum dynamics of geometry, hopefully governed by some form of General Relativistic action, in the continuum limit. For this one has to study either the coarse graining of the lattice path integral appearing \cite{biancacoarse} in our amplitudes, or the renormalization flow and critical behaviour of our GFT model \cite{GFTrenorm}. Also in this respect, our result offer a new, promising concrete model to analyze.

\section*{Acknowledgements}
Support from the `Triangle de la Physique' (Palaiseau-Orsay-Saclay) and the A. von Humboldt Stiftung through a Sofja Kovalevskaja Prize is acknowledged. We also thank V. Bonzom, F. Girelli, F. Hellman, E. Livine, R. Pereira, J. Ryan for useful comments and suggestions.

\end{document}